\documentclass[aps,prb,twocolumn,amsmath,amssymb]{revtex4}
\usepackage{amssymb}

\usepackage{hyperref}
\usepackage{graphicx}
\usepackage{color}

\renewcommand{\a}{\hat{a}} 
\renewcommand{\b}{\hat{b}} 
\renewcommand{\S}{\hat{S}} 
\renewcommand{\Re}{{\rm Re}}

\begin{document}

\title{Static and Dynamic properties of a Fermi-gas of cooled atoms near a
wide Feshbach resonance.}
\author{Deqiang~Sun}
\author{Ar.~Abanov}
\author{V.L.~Pokrovsky}
\affiliation{Department of Physics, MS 4242, Texas A\&M University, College Station, TX
77843-4242 }
\date{\today }

\begin{abstract}
We introduce the Global Spin Model to study the static and dynamic
properties of the ultracold fermionic gas near the broad Feshbach resonance.
We show that the problem of molecular production, in a single-mode
approximation, is reduced to the linear Landau-Zener problem for operators.
The strong interaction leads to significant renormalization of the gap
between adiabatic levels. In contrast to static problem the close vicinity
of exact resonance does not play substantial role. Two main physical results
of our theory is the high sensitivity of molecular production to the initial
value of magnetic field and generation of a large BCS condensate distributed
over a broad range of momenta in a wide range of parameters. We calculate
the amplitude of the condensate as function of the initial detuning and the
rate of the magnetic field sweep.
\end{abstract}

\maketitle

%!!

%\email[]{abanov@tamu.edu}
%\homepage[]{}

%\date{Last Change \LastChange}

%\pacs{PACS numbers:75.10.Nr, 05.50.+q, 75.10.Jm}

\section{Introduction}

\label{sec:intro} In recent years there have been numerous achievements in
the area of ultra-cold atomic physics. The major experimental tool for it is
the use of the Feshbach resonances (FR) \cite{Stwalley1976,Tiesinga1993,
Inouye1998,Mies2000,Goral2004,Chwedenczuk2004,Timmermans1999}, which occurs
when the energy of a quasibound molecular state becomes equal to the energy
of two free alkali atoms. The magnetic-field dependence of the energy allows
precise tuning of the atom-atom interaction strength in an ultracold gas
\cite{Stwalley1976}. Moreover, time-dependent magnetic fields can be used to
reversibly convert atom pairs into weakly bound molecules \cite%
{Regal2003,Strecker2003,Cubizolles2003,Jochim2003,Donley2002,
Chin2003,Herbig2003,Durr2004}. This technique has proved to be extremely
effective in converting degenerate atomic gases of fermions \cite%
{Regal2003,Strecker2003,Cubizolles2003,Hodby2005,Greiner2003,
Jochim2003,Zwierlein2003} and bosons \cite{Claussen2002,Herbig2003,Durr2004}
into bosonic dimer molecules.

 The main theme of this paper is the crossover from the BCS to
BEC condensates. The weakly and strongly paired fermionic s-wave superfluids
were well understood by 60's, while the relation between the superfluids
were studied later by Eagles \cite{Eagles1969}, Leggett \cite{Leggett1980},
and Nozi\'{e}res and Schmitt-Rink \cite{Nozieres1985}. These works treated
the BCS state as a variational ground state and explicitly showed that there
is no qualitative distinction between the BCS and BEC superfluids. After the
discovery of High-$T_{c}$ materials a possibility to describe the pseudogap
phase as a long crossover between the BCS and BEC condensates renewed the
interest to the phenomena \cite{deMelo1993,Levin2005}. It also become very
important in the area of ultracold atomic gases where the
transition/crossover can be directly observed in the molecular production
experiments.

Theoretical works on the molecular production can be roughly referred to two
categories. The first is a phenomenology suggesting that pairs of molecules
independently undergo Landau-Zener (LZ) transitions\cite%
{Mies2000,Chwedenczuk2004}. The total number of molecules at the end of the
process in these works is the LZ transition probability multiplied by the
number of pairs. The most problematic issue in this approach is how to
identify a pair in the gas of indistinguishable fermions. Direct calculation
of the transition probability from a microscopic Hamiltonian, to the 4-th
order in the interaction constant\cite{DP}, shows that, in contrast to the
assumption of phenomenological works, the many-body effects are essential.
Another category includes works based on a simplified model \cite%
{Timmermanns2001} in which molecules have one available state mimicking the
condensate \cite{Tikhonenkov2006-OC,Javanainen2004,Williams2004}. Although
numerical works in this category display a reasonable temperature
dependence, they give no clear physical picture and detailed parameter
dependence.

In this article we consider a homogeneous gas in a box with the large volume
$V$. The gas consists of Fermi isotopes of alkali atoms such as $^{6}$Li, $%
^{24}$Na, $^{40}$K. The gas is subject to an external magnetic field such
that the electronic Zeeman energy strongly exceeds the temperature, whereas
the nuclear Zeeman energy remains much smaller than the temperature.
Therefore, the electronic spins of atoms are almost completely polarized
along the magnetic field. The nuclear spins in the ground state are then
aligned with respect to the electronic spin by the hyperfine interaction. In
a typical experiment \cite{Regal2003,Greiner2003} the magnetic field passes
the Feshbach resonance.
%at which the Zeeman energy of a pair of Fermi-atoms becomes approximately
%equal to the binding energy of the two-atomic molecule.

The origin of this resonance is the following. When two atoms with one
electron in the outer shell each turn into a molecule, the two electrons
%!? "must" is removed
form a spin singlet. This requires one of the electrons to flip its spin,
which costs the Zeeman energy for the electron. The resonance happens when
%!? "This" is removed
Zeeman energy equals to the binding energy of the molecule. Neither exchange
nor magnetic dipolar interaction of electrons can produce such a change of
the spin state. This process is mediated by the hyperfine interaction. The
characteristic scale of the hyperfine interaction $\varepsilon_{hf}$ is in
the range of mK and it is smaller or of the same order of magnitude as the
molecular binding energy. It strongly exceeds the Fermi-energy of the gas
which is in the range of $10^{-6}K$. Collisions in the s--wave channel of
Fermi-atoms is allowed by the Pauli principle only if they have different
states of nuclear spins. Therefore, experimenters create an admixture of
about equal numbers of the fermionic atoms with different nuclear states,
for example atoms of $^{40}$K with the same total atomic spin $9/2$ but
different spin projection quantum numbers $-7/2$ and $-9/2$. Further we will
describe this different states by a pseudospin index $\sigma $ accepting two
values $\uparrow ,\downarrow$.

The difference between the Zeeman energy and its resonance value is called
detuning $\epsilon$. It is the experimentally controlled parameter, which
can be changed by changing external magnetic field. There are two different
types of experimental procedures. In the first one, which we call static, an
equilibrium or metastable state is established at a fixed value of $\epsilon$
and the number of atoms and molecules as well as the distributions of their
velocities as function of detuning are studied. %!? "then" is removed
In the second one, which we call dynamic, the magnetic field sweeps passing
the FR at which the detuning energy $\epsilon $ is zero. After passing
through the resonance a part of atoms are transformed into diatomic bosonic
molecules. The experimental and theoretical problem is the description of
this process and finding the molecular production as a function of the
magnetic field rate. %!? "change" is removed

A major difference between these experiments  is in time
scales. At the first, static, type of experiment the system is always at
thermal equilibrium (quasiequilibrium), while at the second, dynamic, type
the gas is out of equilibrium -- the equilibrating processes are too slow.

A general feature of the equilibrium and dynamic state is the appearance of
the condensates: the Bardeen-Cooper-Schrieffer (BCS) condensate on atomic
side of the resonance and the Bose-Einstein (BE) condensate on its molecular
side. The condensates have the same symmetry. Therefore, the static
transformation from the BCS to BE condensate proceeds not as a sharp
transition, but rather as an eventual crossover.

The main subject of our paper is the BCS and BEC condensates densities and
their correlations in both static and dynamic problems at condition of a
broad resonance which will be shown to be equivalent to a strong
interaction. As always the strong interaction problem is difficult to solve
exactly. We propose a reasonable approximation which allows the analytical
solutions of both, static and dynamic problems.

Below we introduce the necessary notations and definitions.
%following Gurarie and Radzihovsky\cite{Gurarie2007}
The dependence of the s-scattering length $a$ on external magnetic field $B$
is determined by the well-known equation:
\begin{equation}
a\left( B\right) =a_{0}\left( 1-\frac{B_{1}}{B-B_{0}}\right) ,
\label{eq:sclength}
\end{equation}%
where $a_{0}$ is the scattering length far from the resonance; $B_{0}$ is
the resonance value of the field and $B_{1}$ is the magnetic-field width of
the resonance which appear in experimental measurements. An intrinsic energy
scale generated by the field width is \cite{Gurarie2007}:%
\begin{equation}
\tilde{\Delta}=\frac{4\mu _{B}^{2}B_{1}^{2}ma_{0}^{2}}{\hbar ^{2}}
\label{Delta-tilde}
\end{equation}%
We will show later that the value of $\tilde{\Delta}$ characterises the
interaction in the Fermi-gas or the interaction of the BCS pairs with the
BEC condensate.

There are two different regimes for both static and dynamic transitions:
narrow and broad Feshbach resonance. The Feshbach resonance is narrow when $%
\tilde{\Delta}$ is much smaller than the Fermi energy $\varepsilon _{F}$ and
broad in the opposite case. Thus, the resonance is broad if the
dimensionless parameter%
\begin{equation}
\Gamma =\sqrt{\frac{\tilde{\Delta}}{\varepsilon _{F}}}  \label{Gamma}
\end{equation}%
is large. In the opposite case the resonance is narrow. The static and
dynamics of the narrow resonance was thoroughly analyzed theoretically by
Gurarie and Radzihovsky to whose review \cite{Gurarie2007} we refer the
reader. The case of the broad resonance was not analyzed since it is a
strong interaction problem. We show that the condition of the broad
resonance \eqref{eq:broad} allows simplification of the model despite of
strong interaction between fermions. The key idea is a proper cut-off in the
momentum space and neglect of the fermion dispersion. The resulting model is
similar to the Dicke spin model for superradiance \cite{dicke}. This model
allows us to solve the static problem exactly. The complete spectrum and
eigenstates are found. The solution displays a crossover from BCS to BEC in
a range of detuning near the FR. We find the density of the condensates and
their correlation as function of the detuning. We show that, contrary to the
standard BCS solution, the scale of momenta in which the Cooper pair wave
function is confined is $(2m\tilde{\Delta})^{1/2}$, much larger than the
Fermi energy. Therefore, the size of the Cooper pair in real space is much
less than the average distance between atoms. We discuss consequences of
this fact for both statics and dynamics.

The dynamic problem is reduced to the Landau-Zener problem in the operator
form which also can be solved exactly. One of conservation laws in our
theory ensures that the increase of one of condensates proceeds only at
expense of another. The main conclusions of the dynamic problem are: i)
Sensitivity of the molecular production to the initial value of the BCS
condensate; ii) Conservation of a sum of weighted BCS and BEC condensate
densities which indicates that one of condensates can grow only at the
expense of another one. %!? the next sentence is omitted.
The dynamic problem was briefly described earlier in our article \cite{SAP}.

\section{ The model and conservation laws}

\label{sec:broad}

We start with the Timmermans \cite{Timmermans1999} Hamiltonian \footnote{%
For simplicity, we focus on a homogeneous system of volume $V$}:
\begin{eqnarray}
&&\hat{H}=\sum_{\mathbf{p},\sigma }(\epsilon +\varepsilon _{\mathbf{p}})\hat{%
a}_{\mathbf{p}\sigma }^{\dagger }\hat{a}_{\mathbf{p}\sigma }+\sum_{p}\omega
_{\mathbf{q}}\hat{b}_{\mathbf{q}}^{\dagger }\hat{b}_{\mathbf{q}}+  \nonumber
\\
&&\frac{g}{\sqrt{V}}\sum_{\mathbf{p,q}}\left( \hat{b}_{\mathbf{q}}\hat{a}_{%
\mathbf{p+q}\uparrow }^{\dagger }\hat{a}_{-\mathbf{p}\downarrow }^{\dagger }+%
\hat{b}_{\mathbf{q}}^{\dagger }\hat{a}_{-\mathbf{p}\downarrow }\hat{a}_{%
\mathbf{p+q}\uparrow }\right)  \label{eq:Ham}
\end{eqnarray}%
Here $\hat{a}_{p\sigma }^{\dagger }$ are creation operators of fermionic
atoms with momentum $\mathbf{p}$, pseudospin $\sigma $, and kinetic energy $%
\varepsilon _{\mathbf{p}}=\mathbf{p}^{2}/2m$; $\hat{b}_{\mathbf{q}}^{\dagger
}$ are the creation operators of the bosonic diatomic molecule with kinetic
energy $\omega _{\mathbf{q}}=\mathbf{q}^{2}/4m$; the position of the FR is
controlled by the experimentally tunable detuning energy $\epsilon $, wich
becomes time-dependent in the dynamic problem. The last term in %
\eqref{eq:Ham} is the interaction with a coupling constant $g$. It describes
the formation of a molecule from two atoms and the inverse process of
dissociation into two atoms. %; the coupling constant $g$ relates to
%the reaction of molecule formation from two atoms and the inverse process
%of molecular dissociation.

As it was explained earlier, the coupling constant $g$ is provided by the
hyperfine interaction. Its value can be estimated \cite{LLv3} as $g\sim
\varepsilon _{hf}\sqrt{a_{m}^{3}}$, where $\varepsilon _{hf}$ is a
characteristic hyperfine energy (about 1mK) and $a_{m}$ is the size of the
diatomic molecule. The Hamiltonian \eqref{eq:Ham} neglects nonresonant
direct atom-atom and molecule-molecule interactions that near any FR are
subdominant to the resonant scattering. To some extent the direct atom-atom
interaction is taken into account by the value $a_{0}$ in eq. (\ref%
{eq:sclength}). We use the word ``atoms'' for both uncoupled atoms and atoms
within molecules, and the word ``fermions'' for atoms that are not bound in
molecules. Correspondingly, we denote the number of atoms as $N$ and the
number of fermions as $\hat{N}_{F}=\sum_{\mathbf{p},\sigma }\hat{a}%
^{\dagger}_{\mathbf{p}, \sigma}\a_{\mathbf{p},\sigma } $.  Since the
diatomic molecule can be created by the interaction term in \eqref{eq:Ham}
only at expense of two fermions the Hamiltonian (\ref{eq:Ham}) conserves the
total number of atoms:%
\begin{equation}
N=\hat{N}_{F}+2\sum_{\mathbf{q}}\hat{b}_{\mathbf{q}}^{\dagger }\hat{b}_{%
\mathbf{q}}  \label{eq:atom-number}
\end{equation}%
Therefore, in the Hamiltonian (\ref{eq:Ham}) the detuning energy can be
transferred with the coefficient $-2$ to the energy of molecules. To find
the connection between the coupling constant $g$ in the Hamiltonian (\ref%
{eq:Ham}) and the value $\tilde{\Delta}$ defined by equation (\ref%
{Delta-tilde}) we use the perturbation theory to the second order to
eliminate $\hat{b}_{\mathbf{q}}$ and $\hat{b}_{\mathbf{q}}^{\dagger }$ from
the static Hamiltonian (\ref{eq:Ham}) and obtain the Hamiltonian for
fermions only with 4-fermion interaction. Assuming that $\varepsilon _{%
\mathbf{p}}$ and $\omega _{\mathbf{q}}$ can be neglected in comparison to $%
\epsilon $ (this assumption will be justified later), the interaction
Hamiltonian reads:
\begin{equation}
H_{int}=\frac{g^{2}}{2V\epsilon }\sum_{\mathbf{p,p}^{\prime },\mathbf{q}}%
\hat{a}_{\mathbf{p+q}\uparrow }^{\dagger }\hat{a}_{-\mathbf{p}\downarrow
}^{\dagger }\hat{a}_{-\mathbf{p}^{\prime }\downarrow }\hat{a}_{\mathbf{p}%
^{\prime }\mathbf{+q}\uparrow }  \label{eq:Ham-eff}
\end{equation}%
Thus, the effective interaction constant for fermions is $%
g_{F}=g^{2}/2\epsilon $. It is negative (attraction) at negative $\epsilon $%
. The s-scattering length $a$ is related to the interaction constant $g_{F}$
by a standard relation\cite{LLv9} $g_{F}=4\pi \hbar ^{2}a/m$. Thus, the
singular part of the scattering length is associated with the fermion-boson
coupling constant $g$ and the detuning energy $\epsilon $ as follows:
\begin{equation}
a(\epsilon )=\frac{mg^{2}}{8\pi \hbar ^{2}\epsilon }
\label{eq:s-length-theor}
\end{equation}%
Comparing equation (\ref{eq:s-length-theor}) with the resonance term in
equation \eqref{eq:sclength} and identifying $\epsilon =\mu
_{B}\left(B-B_{0}\right)$, we arrive at a relation $2B_{1}a_{0}%
\mu_{B}=mg^{2}/4\pi \hbar^{2}$. Using $\epsilon _{F}=\hbar^{2}(3\pi^2 n
)^{2/3}/2m$, where $n$ is the density of fermions, and substituting this
into equations (\ref{Delta-tilde},\ref{Gamma}) we express the intrinsic
energy $\tilde{\Delta}$ and dimensionless parameter $\Gamma $ in terms of
the coupling constant $g$:
\begin{eqnarray}
&&\tilde{\Delta}=\frac{m^{3}g^{4}}{16\pi ^{2}\hbar ^{6}}  \label{Delta-g} \\
&&\Gamma=\frac{m^{2}g^{2}}{\hbar ^{4}n^{1/3}}\frac{1}{\pi^{5/3}3^{1/3}
2^{3/2}}  \label{Gamma-g}
\end{eqnarray}
The Hamiltonian (\ref{eq:Ham}) is still too complicated. We show that the
broad resonance condition
\begin{equation}  \label{eq:broad}
\Gamma \gg 1
\end{equation}
allows us to apply two simplifying assumptions.

The first one of them is the single-mode approximation: we neglect all
bosonic modes except the one with zero momentum. The characteristic energy
scale in statics and dynamics is $\tilde{\Delta}$. The corresponding value
of the scattering length following from equation \eqref{Delta-g} at such a
detuning is $a(\tilde{\Delta })\sim n^{-1/3}/\Gamma$, which gives the
characteristic value of the gas parameter $an^{1/3}\sim \Gamma^{-1} \ll 1$.
Small $|\epsilon| \sim \tilde{\Delta}/\Gamma$ correspond to the strong
interaction regime in which the fermions are distributed in a broad range of
momentum $\sim \sqrt{2m\tilde{\Delta}}$. The maximal modulus of the
scattering amplitude $f$ for such particles achieved in the unitary limit is
$f\sim \hbar /\sqrt{2m\tilde{\Delta}}$. The corresponding value of the gas
parameter $fn^{1/3}$ is again of the order $\Gamma^{-1} \ll 1$. At this
conditions and at zero temperature almost all particles of the Bose-gas fall
into the coherent condensate with zero momentum. We thus will substitute $%
\hat{b}_{\mathbf{q}}=\delta_{0,\mathbf{q}}b_{0}$ in the Hamiltonian %
\eqref{eq:Ham}. %!? Here I replaced 3 dashes by the dot.
This is the single-mode approximation.

The applicability of the single-mode approximation to the dynamics seems
less obvious since the formation of the condensate in the Bose gas requires
some time which must be compared with the characteristic time of the
fermion-boson transition. The kinetic of the condensate formation was
theoretically studied by Kagan \textit{et al.} \cite{Kagan-Svist-Shlyap,
Kagan-Svist} and by Gardiner \textit{et al.} \cite{Gardiner1}. In the
article \cite{Kagan-Svist} it was concluded that the formation of condensate
starts after some cooling process which requires a delay time, typically few
tens of ms in the laser cooled gases and then the condensate grows rapidly.
In the experiment \cite{Gardiner2} even smaller delay time was found. The
characteristic time of sweeping through the FR was of the same order of
magnitude or longer. Besides, the experimenters started the sweeping in a
close vicinity of the resonance in which at least the BCS condensate was
sufficiently strong. As we will see later, in dynamics it produces the BEC
condensate during the time characteristic for the Landau-Zener processes.

The second approximation we apply is disregarding the fermionic dispersion
in \eqref{eq:Ham} in comparison to $\varepsilon $. A typical value of $b_{0}$
is $\sqrt{N}$. In the broad resonance approximation $g\sqrt{N}/\sqrt{V}=g%
\sqrt{n}$ is much larger than $\varepsilon _{F}\sim \hbar ^{2}n^{2/3}/m$, $gm%
\sqrt{n}/\hbar^{2}n^{2/3} \sim \sqrt{\Gamma }\gg 1$, see \eqref{Gamma-g}.
With this motivation we neglect the kinetic energy of fermions in comparison
to $\varepsilon $. Though rather rough, this approximation retain essential
features of the exact Hamiltonian. The character and precision of this
approximation will be discussed in more details later.

These two approximations  greatly simplify the Hamiltonian %
\eqref{eq:Ham} reducing it to the following form:
\begin{eqnarray}
&& H_{sm}=\sum_{\mathbf{p},\sigma }^{p_{s}}\epsilon\hat{a}^{\dagger}_{%
\mathbf{p}\sigma } \a_{\mathbf{p}\sigma }+  \label{Ham-s-mode} \\
&& \frac{g}{\sqrt{V}}\left[ b_{0}\sum_{\mathbf{p}}^{p_{s}}a_{\mathbf{p}%
\uparrow }^{\dag }a_{-\mathbf{p}\downarrow }^{\dag }+ b_{0}^{\dag }\sum_{%
\mathbf{p}}^{p_{s}}a_{\mathbf{p}\downarrow }a_{-\mathbf{p}\uparrow }\right]
\nonumber
\end{eqnarray}
The number of atoms conserved by the Hamiltonian \eqref{Ham-s-mode} reads:
\begin{equation}
N=\sum_{\mathbf{p},\sigma }^{p_{s}}\hat{a}^{\dagger}_{\mathbf{p}\sigma } \a_{%
\mathbf{p}\sigma }+2b_{0}^{\dag }b_{0}  \label{eq:atom-number-s}
\end{equation}
In Eqs. \eqref{Ham-s-mode} and \eqref{eq:atom-number-s} we have introduced
the cutoff momentum $p_{s}$ to account for the dropped fermionic dispersion.
The value of this momentum must be such that our approximation of neglecting
the kinetic energy in comparison to $\tilde{\Delta}$ is %would be
justified. Thus, a reasonable cut-off momentum is:
\begin{equation}
p_{s}=\sqrt{2m\tilde{\Delta}}  \label{ps-Delta}
\end{equation}%
Plugging eq. \eqref{Delta-g} into eq. \eqref{ps-Delta}, we find:%
\begin{equation}
p_{s}=\frac{m^{2}g^{2}}{2^{3/2}\pi \hbar ^{3}}  \label{ps-g}
\end{equation}

The number of available states $N_s$, which includes all the states with $%
p<p_{s}$, reads:
\begin{equation}
N_{s}=V\frac{p_{s}^{3}}{3\pi ^{2}\hslash ^{3}}  \label{Ns-ps}
\end{equation}
Eq. \eqref{ps-g} gives the following result for the density of the available
states $n_{s}=N_{s}/V$:
\begin{equation}
n_{s}=\frac{m^{6}g^{6}}{2^{9/2}\cdot 3\pi ^{5}\hslash ^{12}}  \label{ns-g}
\end{equation}%
With precision of the coefficient 1.09 the value
\begin{equation}  \label{eq:D}
\Delta =g\sqrt{n_{s}}
\end{equation}
coincides with $\tilde{\Delta}$. The condition of the broad resonance is
equivalent to the strong inequality:%
\begin{equation}
N_{s}\gg N  \label{NsggN}
\end{equation}

Let us consider the whole Hilbert space as a set of pairs of conjugated
states $|\mathbf{p},\uparrow \rangle$ and $|{-\mathbf{p}},\downarrow \rangle$%
. Each pair can be either empty, or single occupied, or doubly occupied. A
single occupied pair is not changed by the Hamiltonian \eqref{Ham-s-mode}.
The whole Hilbert space can then be split on two invariant subspaces: the
set of the single occupied pairs and the rest, where no state is single
occupied. The Hamiltonian is diagonal in the first subspace, so it induces
no transitions in it. We thus consider only an invariant subspace of the
Hilbert space with all pairs either empty or doubly occupied.

Besides of the occupation number of fermionic pairs, the state is fixed with
the number of bosonic molecules $N_m$. The Hilbert subspace invariant under
the action of the Hamiltonian \eqref{Ham-s-mode} at a fixed total value of
atoms $N$ consists of states with given number $N_m$ of molecules in the
condensate and the rest of atoms occupying pairs of fermionic states. The
number of occupied pairs is $M=N/2-N_{m}$. Summing the number of states with
different possible $M$, we find for the dimensionality of the Hilbert space $%
\mathcal{D}( N,N_{s})$ with given $N$ and $N_{s}$:
\begin{equation}
\mathcal{D}( N,N_{s}) =\sum_{M=0}^{N/2}\frac{N_{s}!}{ M!( N_{s}-M) !}
\label{Hilbert-D}
\end{equation}

\section{The global spin model and its Hilbert space}

Following Anderson \cite{AndersonTrans}, we introduce spin operators:
\begin{eqnarray}
&&s_{\mathbf{p}z} =\frac{1}{2}\left( a_{\mathbf{p\uparrow }}^{\dag }a_{%
\mathbf{p\uparrow }}+a_{-\mathbf{p\downarrow }}^{\dag }a_{-\mathbf{%
p\downarrow }}-1\right)  \label{sz} \\
&&s_{\mathbf{p}+} =a_{-\mathbf{p\downarrow }}^{\dag }a_{\mathbf{p\uparrow }%
}^{\dag }; \qquad s_{\mathbf{p-}}=a_{\mathbf{p\uparrow }}a_{-\mathbf{%
p\downarrow }}  \label{s+}
\end{eqnarray}
In the double-occupied or empty fermionic pairs subspace they obey the
standard spin-1/2 commutation relations:
\begin{equation}
\left[ s_{\mathbf{p}z},s_{\mathbf{p}^{\prime }\pm }\right] =\pm \delta _{%
\mathbf{pp}^{\prime }}s_{\mathbf{p}\pm };~\left[ s_{\mathbf{p}+},s_{\mathbf{p%
}^{\prime }\mathbf{-}}\right] =2\delta _{\mathbf{pp}^{\prime }}s_{\mathbf{p}%
z}  \label{commutation}
\end{equation}
so that the double occupied pair corresponds to $s_{\mathbf{p}z}=+\frac{1}{2}
$ and empty pair corresponds to $s_{\mathbf{p}z}=-\frac{1}{2}$. Note that a
single-occupied fermionic pair corresponds to a singlet spin state: all
three spin operators (\ref{sz}, \ref{s+}) turn such a pair to zero.

The neglect of kinetic energy allows us to rewrite the Hamiltonian %
\eqref{Ham-s-mode} in terms of only global spin operators:
\begin{equation}
H_{S}=2\epsilon S_{z}+\frac{g}{\sqrt{V}}\left( b_{0}S_{+}+b_{0}^{\dag
}S_{-}\right)  \label{Ham-spin-global}
\end{equation}%
(we have omitted a constant originated from the term $-1$ in equation %
\eqref{sz}), where the global spin components
\begin{equation}
S_{z}=\sum s_{\mathbf{p}z};\qquad S_{\pm }=\sum s_{\mathbf{p}\pm }
\label{spin-global}
\end{equation}%
obey the standard commutation relationships: $\left[ S_{z},S_{\pm }\right]
=\pm S_{\pm };~\left[ S_{+,}S_{-}\right] =2S_{z}$. We will call the model
with the Hamiltonian \eqref{Ham-spin-global} the Global Spin Model (GSM).

A subtle assumption incorporated in the derivation of the Hamiltonians %
\eqref{Ham-s-mode}, \eqref{Ham-spin-global}, and in the permutation
relations for the global spin components is that the summations in the two
definitions \eqref{spin-global} run over the same range of momenta. It is
not obvious since the summation in the sum for $S_{z}$ is limited by the
condition $\varepsilon _{\mathbf{p}} < \left\vert\epsilon \right\vert $, %!! < instead of \ll
whereas the summation in the second sum originated from the boson-fermion
interaction part of the Hamiltonian is naturally cut off by the range of
interaction. It means that $g$ is a function of momentum vanishing at
sufficiently large values of the momentum modulus. In  the dynamical problem the
Anderson spins rotate with the frequencies $\left(\epsilon +\varepsilon _{%
\mathbf{p}}\right) /\hslash $. Therefore they rotate coherently and enhance
remarkably their effective field exerted to the moleculare amplitude $b_{0}$
only if $\varepsilon _{\mathbf{p}}<\left\vert \epsilon \right\vert $. We
will see that such a coherence indeed takes place in the dynamic problem.
The contributions from larger values of momenta to $S_{\pm }$ are incoherent
and mutually compensate their effect. Thus, for dynamic problem, the
summation in the same momentum region in the two equations %
\eqref{spin-global} is justified.

It is not so obvious for the static problem. We will show that the static
GSM has many qualitative features resembling what is expected for the
initial Timmermanns\cite{Timmermans1999} \textit{et al.} model and observed
in the experiment, though nobody solved the latter model exactly. In
particular the GSM displays a crossover from BCS to BEC condensate with a
large gap due to strong interaction in a broad vicinity of the Feshbach
resonance. However, in quantitative details they are different. Besides, the
GSM \eqref{Ham-spin-global} has an additional symmetry and an additional
conserved value which the Timmermans Hamiltonian does not possess. Indeed
the Hamiltonian \eqref{Ham-spin-global} conserves not only the value of
\begin{equation}
Q=S_{z}+b_{0}^{\dag }b_{0},  \label{Sz-conservation}
\end{equation}
equivalent to the number of atoms \eqref{eq:atom-number-s}, but also the
total spin $S$, where:
\begin{equation}
S\left( S+1\right) =S_{z}^{2}+\frac{1}{2}\left( S_{+}S_{\_}+S_{-}S_{+}\right)
\label{total-spin}
\end{equation}%
In the thermodynamic limit of large system it is possible to neglect $1$ in
comparison to $S$ and consider $S_{+}$ and $S_{-}$ as commutative values.
The conservation laws (\ref{total-spin}, \ref{Sz-conservation}) imply that
the quantity of one of the condensates (BCS or BEC) can be increased only at
the expense of the other.

The cut-off in the momentum space define a finite-dimensional Hilbert space
of states $R_N$. What we call the GSM is this Hilbert space and the
Hamiltonian (\ref{Ham-spin-global}) acting in it.

Each state in $R_N$ is a vector from a direct product of $N_{s}$ spin 1/2
representations corresponding to the Anderson spins $\mathbf{s}_{\mathbf{p}}$%
. The combinatorial coefficient $N_{s}!/M!(N_{s}-M)!$ entering eq. %
\eqref{Hilbert-D} can be treated as the number of possible distributions of $%
M$ spins up and $N_{s}-M$ spins down, i.e. the number of different states
with the spin projection:
\begin{equation}
S_{z}=M-\frac{N_{s}}{2}  \label{Sz-M-Ns}
\end{equation}
Since $0\leq M\leq N/2$, the projection $S_z$ at a fixed $S$ is limited by      %!! I removed "for $S_z$"
inequalities: 
\begin{equation}
-S\leq S_{z}\leq (N-N_{s})/2.                     %!! This inequality is separated to an equation with number to reference it further
\label{Sz-range}
\end{equation}
According to \eqref{NsggN}, $S_{z}$ is always large and negative. The total spin $S$
cannot be smaller than $|S_{z}|$, it also cannot be larger than $N_{s}/2$.
Thus, the allowed values of the total spin $S$ are $(N_{s}-N)/2 \leq S \leq
N_{s}/2$.

The number $\mathcal{N}(S,N_s)$ of different multiplets with a fixed value
of the total spin $S$ which appear at the addition of $N_s$ spins 1/2 reads
\cite{LLv3} %\cite{Lifshitzbook}:
\begin{equation}
\mathcal{N}\left( S, N_s\right) =\frac{N_{s}!\left( 2S+1\right) }{\left(
\frac{N_{s}}{2}-S+1\right) !\left( \frac{N_{s}}{2}+S\right) !}  \label{N-S}
\end{equation}%
Each of these representations contains generally $2S+1$ states, but only $S-%
\frac{N_{s}-N}{2}$ of them are allowed by the inequalities \eqref{Sz-range}. The number
of states $\tilde{\mathcal{N}}(S,S_z)$ in the Hilbert space of the GSM with
fixed values $S$ and $S_z$ reads:
\begin{equation}
\tilde{\mathcal{N}}(S,S_z) =\theta (S_z+S)\theta \left(-\frac{N_s}{2}+\frac{N%
}{2}-S_z\right) \mathcal{N}(S),  \label{N-S-Sz}
\end{equation}
where $\theta (x)$ is the Heaviside step function.

The Hamiltonian \eqref{Ham-spin-global} formally coincides with the famous
Dicke Hamiltonian for the so-called superradiance problem \cite{dicke}, but
it acts in a different Hilbert space.

\section{Spectra and eigenstates of the GSM}

The Hamiltonian \eqref{Ham-spin-global} commutes with $S$ and $N$.
Therefore, its stationary states $|\Psi\rangle $ at fixed values $N_{s}$, $N$%
, and $S$ can be represented by a superposition of states with fixed number
of molecules $N_{m}$ or $S_{z}=\frac{N-N_{s}}{2}-N_{m}$:
\begin{equation}
\left\vert \Psi \right\rangle =\sum_{N_{m}=0}^{\frac{N-N_{s}}{2}+S}\Psi
_{N_{m}}\left\vert N_{m}\right\rangle ,  \label{superposition}
\end{equation}%
whose amplitudes $\Psi _{N_{m}}$ obey the stationary Schr\"{o}dinger
equation:
\begin{eqnarray}
&&E\Psi _{N_{m}}=-2\epsilon N_{m}\Psi _{N_{m}}+  \nonumber \\
&&\frac{g}{\sqrt{V}}\sqrt{N_{m}\left( S-S_{z}\right) \left( S+S_{z}+1\right)
}\Psi _{N_{m}-1}+ \\
&&\frac{g}{\sqrt{V}}\sqrt{\left( N_{m}+1\right) \left( S+S_{z}\right) \left(
S-S_{z}+1\right) }\Psi _{N_{m}+1}  \nonumber  \label{SE}
\end{eqnarray}
This system is still rather complicated, but it is strongly simplified by
the broad resonance condition \eqref{NsggN}. Indeed, due to this inequality $%
S\approx -S_{z}\approx N_{s}/2$, and it is possible to replace $S-S_{z}$ and
$S-S_{z}+1$ in eq. \eqref{SE} by $N_{s}$. We should be more careful with the
expression $S+S_{z}$ since the two terms almost completely cancel each
other. Employing the notation defined earlier $\Delta =g\sqrt{n_{s}}$ %
\eqref{eq:D} and introducing new variables
\begin{equation}
S_{z}=s-S-m;\quad s=\frac{S}{2}+\frac{N-N_{s}}{4};\quad m=N_{m}-s,
\label{reduced-variables}
\end{equation}%
we arrive at a simplified version of equations (\ref{SE}):%
\begin{eqnarray}
&&-2\epsilon m\Psi _{m}+\Delta \sqrt{(s-m)(s+m+1) }\Psi _{m+1}+  \nonumber \\
&&\Delta \sqrt{( s+m)( s-m+1) }\Psi_{m+1} = ( E+2s\epsilon) \Psi _{m},
\label{reduced-equation}
\end{eqnarray}
Since $N_{m}\geq 0$ and $S+S_{z}\geq 0$, the quantum number $m$ runs from $-s
$ to $s$ and $0\leq s \leq N/4$. The equation \eqref{reduced-equation} is
easily recognizable as generated by a reduced spin Hamiltonian:
\begin{equation}
H_{r}=-2\epsilon s_{z}+2\Delta s_{x}  \label{reduced-Hamiltonian}
\end{equation}%
where $s_{x}$, $s_{z}$ are spin operators corresponding to the reduced total
spin $s$. This is a Hamiltonian of a spin $s$ in the magnetic field $2\sqrt{%
\epsilon^{2}+\Delta^{2}}$ tilted in the $xz$ plane at the angle $\theta
=-\tan^{-1}(\Delta /\epsilon) $ to the $z$ axis.

The energy levels are labeled by two integers $s$ and $\tilde{m}$ (we have
introduced tilde to distinguish projection to the axis tilted by the angle $%
\theta$ to $z-$axis from projection to the $z-$axis):
\begin{equation}
E_{s\tilde{m}}\left( \epsilon \right) =2\tilde{m}\sqrt{\epsilon ^{2}+\Delta
^{2}}-2s\epsilon  \label{spectrum}
\end{equation}%
The spectrum (\ref{spectrum}) possesses a symmetry:
\begin{equation}
E_{s\tilde{m}}\left( \epsilon \right) =-E_{s,-\tilde{m}}\left( -\epsilon
\right)  \label{symmetry}
\end{equation}%
Levels with the same $s$ and different $\tilde{m}$ do not cross, but the
levels with different $s$ and $\tilde{m}$ cross each other. The crossing of
the levels $\left( s,\tilde{m}\right) $ and $\left( s^{\prime },\tilde{m}%
^{\prime }\right) $ happens at a point:%
\begin{equation}
\epsilon =\Delta \frac{\mathrm{sign}\left( \frac{s-s^{\prime }}{\tilde{m}-%
\tilde{m}^{\prime }}\right) }{\sqrt{\left( \frac{s-s^{\prime }}{\tilde{m}-%
\tilde{m}^{\prime }}\right) ^{2}-1}},  \quad |s-s^{\prime}|>|%
\tilde{m}-\tilde{m}^{\prime}|  \label{crossing}
\end{equation}%
Besides of crossings each level $\left( s,\tilde{m}\right) $ at any $%
\epsilon $ is $\mathcal{N}( S,S_{z})$ fold degenerate as it is determined by
eq. \eqref{N-S}. In terms of reduced spin variables it reads:
\begin{equation}
\mathcal{N}(N,s,\tilde{m})\approx \theta (s-\tilde{m})\theta (s+\tilde{m})%
\frac{N_s^{N/2-2s+1}e^{-N/2+2s}}{\sqrt{2\pi}(N/2-2s)!}  \label{N-s-m}
\end{equation}

For a fixed $s$ the state with minimal energy is $( s,-s) $. The ground
state corresponds to the maximal possible value $s=N/4$ and $\tilde{m}=-s$. Its
energy is:
\begin{equation}
E_{G}=E_{N/4,N/4}=-\frac{N}{2}\left( \sqrt{\epsilon ^{2}+\Delta ^{2}}%
+\epsilon \right)  \label{Ground-energy}
\end{equation}%
The energy of the ground state is separated from the rest of spectrum by a
finite energy gap $\delta$, which is determined by the following equation:
\begin{equation}
\delta = 2\left[\sqrt{\epsilon^2+\Delta^2}+\epsilon\theta(-\epsilon )\right].
\end{equation}
The first excited state at negative $\epsilon$ is the state with $s=\frac{N}{%
4}-1$ and $\tilde{m}=-\frac{N}{4}+1$; at positive $\epsilon$ it is the state
with $s=N/4$ and $\tilde{m}=-\frac{N}{4}+1$. In the ground state the maximal
spin $s=N/4$ is oriented in the $xz-$plane at the angle $\theta
=\tan^{-1}(\Delta/\epsilon )$ to the $z-$axis. This state can be thought of
as a set of $N/2$ spins $1/2$, all directed along the same line. The
corresponding wave function is $|\psi\rangle =\prod_{i=1}^{N/2} \otimes
|\psi_{i}\rangle$, where $|\psi_{i}\rangle=\left(\cos \theta /2,\sin \theta
/2 \right)_{i}^{T}$. The operator $s_{z}$ is represented by the direct sum: $%
s_{z}=\sum_{i=1}^{N/2}s_{z}^{i}$. Therefore, the average value of $m$ in the
ground state is equal to $\langle m\rangle _{G}=\langle\psi | s_{z}|\psi
\rangle=\sum_{i=1}^{N/2} \langle \psi_{i}|s_{z}^{i}|\psi_{i}\rangle =\frac{N%
}{4}\cos \theta =N\epsilon /4\sqrt{\epsilon ^{2}+\Delta ^{2}}$. Employing
the third equation \eqref{reduced-variables}, we find the average number of
molecules in the ground state:%
\begin{equation}
\langle N_{m}\rangle _{G}=\frac{N}{4}\left( 1+\frac{\epsilon }{\sqrt{%
\epsilon ^{2}+\Delta ^{2}}}\right)  \label{M-G}
\end{equation}
It smoothly varies from 0 at $\epsilon =-\infty $ to $N/2$ at $\epsilon
=+\infty $. The width of the transition is determined by $\Delta $. The
value $\left\langle N_{m}\right\rangle _{G}-\frac{N}{4}$ is an odd function
of the detuning energy. The average number of fermions $\left\langle
N_{F}\right\rangle $ can be found from the conservation law $N_{F}+2N_m=N$.

The fluctuation of the number of molecules in the ground state also can be
calculated exactly. Indeed, $\langle (\Delta N_{m})^{2}\rangle_{G}=\langle
\psi | (s_{z})^{2}|\psi \rangle- \langle \psi | s_{z}|\psi \rangle^{2}$
using $\langle \psi | (s_{z})^{2}|\psi \rangle=N/8+\sum_{i\not=j}^{N/2}
\langle \psi_{i}|s_{z}^{i}|\psi_{i}\rangle \langle
\psi_{j}|s_{z}^{j}|\psi_{j}\rangle =N/8+\cos^{2} (\theta) N(N-2)/16 $. Thus,
at large $N$, the quadratic fluctuation of the number of molecules reads:
\begin{equation}  \label{fluctuation}
\langle (\Delta N_{m})^{2}\rangle_{G}= \frac{N}{8} \frac{\Delta^{2}}{%
\epsilon^{2}+\Delta^{2}}
\end{equation}
The fluctuation $\langle ( \Delta N_{m}) ^{2}\rangle$ is maximal at $%
\epsilon =0$ --- at the FR, and is an even function of the
detuning $\epsilon $.

An important value is the BCS condensate amplitude $\langle S_{+}\rangle_{G}
$ in the ground state. Using the first equation \eqref{reduced-variables}
and \eqref{total-spin}, we find that $|\langle S_{+}\rangle_{G}|^{2}\approx
S^{2}-\langle S^{2}_{z}\rangle_{G} \approx
(N_{s}/2)^{2}-(N_{s}/2-N/4)^{2}-N_{s}\langle s_{z}\rangle_{G}\approx \frac{1%
}{4}NN_{s}(1-\cos \theta )$ and thus
\begin{equation}  \label{eq:BCS-epsilon}
|\langle S_{+}\rangle_{G}|=\sqrt{\frac{NN_{s}}{2}}|\sin \theta /2|
\end{equation}
The ground state is symmetric with respect to any permutation of the $N_s$
pair states, which is the Hamiltonian symmetry operation. Therefore, all
amplitudes $\langle a^{\dag}_{\mathbf{p}\uparrow}a^{\dag}_{-\mathbf{p}%
\downarrow}\rangle$ are equal and each of them is equal to $\sqrt{N/N_s}%
\sin(\theta/2)$. A simple physical picture behind this distribution is the
following. The total number of occupied pairs is equal to $N\sin^2(\theta/2)$. 
They are equally distributed among $N_s$ available states. Therefore, the
probability to find a pair at any fixed state is $N\sin^2(\theta/2)/N_s$ and
amplitude to find this state is $\sqrt{N/N_s}\sin(\theta/2)\times
e^{i\varphi_{\mathbf{p}}}$. This result coincides with indicated above if
all amplitudes are coherent and have the same phase. According to this
picture only the total number of pairs depend on detuning, but their
distribution in momentum space remains unchanged: they are uniformly
distributed over momenta from zero to $p_s$. It means that the size of the
Cooper pair is always $\hbar /p_s$. it is much smaller than the distance
between gas particles $n^{-1/3}$. The pairs are compact, but still the
collective interaction of the fermions inside the pairs play important role.
The implicit assumption of the theory is that the size of the Bosonic
molecule is much less than $\hbar /p_s$. Another important difference
between the Cooper pair and molecule is that in the former electron spins of
atoms are parallel being polarized by magnetic field, whereas in the latter
the electron spins are antiparallel. The broad distribution of the Cooper
wave function in the momentum space is a consequence of the corresponding
broad range of the hyperfine interaction. This property will persist in a
more accurate theory which does not accept our simplifying approximations.

We notice from \eqref{M-G} and \eqref{eq:BCS-epsilon} that the value
\begin{equation}  \label{eq:staticConservation}
N_{s}N_{m}+|\langle S_{+}\rangle|^{2}=N_{s}N/2
\end{equation}
is a detuning independent constant.

Finally, we find the BEC-BCS correlation function $\left\langle
b_{0}S_{+}\right\rangle $. We demonstrated earlier (see the derivation of
eq. \eqref{reduced-equation} from \eqref{SE}) that in the broad resonance
approximation $N_{s}\gg N$ the product of operators $b_{0}S_{+}$ can be
replaced by $\sqrt{N_{s}}s_{+}$. Thus,
\begin{equation}
\left\langle b_{0}S_{+}\right\rangle =\frac{\sqrt{N_{s}}N}{4}\sin \theta =%
\frac{\sqrt{N_{s}}N}{4}\frac{\Delta }{\sqrt{\epsilon ^{2}+\Delta ^{2}}}
\label{BCS-BEC}
\end{equation}%
This correlator vanishes at $\epsilon =\pm \infty $ and has maximum at $%
\epsilon =0$.   An interesting relationship between the BEC-BCS correlator
and the fluctuation of the number of molecules follows from 
eqns. \eqref{fluctuation} and \eqref{BCS-BEC}: $N_{s}N\langle (\Delta
N_{m})^{2} \rangle_{G}=2\left\langle b_{0}S_{+}\right\rangle^{2}$.

The GSM displays BCS-BEC crossover in the range $\left\vert \epsilon
\right\vert \sim \Delta $ near the Feshbach resonance. In this range the BCS
condensate amplitude grows to the value $\sim \sqrt{N_{s}N/2}\gg N/2$. This
enhancement of the condensate is due to the distribution of the Cooper pairs
over a wide range of momenta strongly exceeding the Fermi sphere. It
indicates that the famous BCS exponentially small condensate does not appear
even at very large detuning exceeding $\Delta $. In the BCS theory the
condensate or energy gap is exponentially small not only due to the weakness
of interaction, but also because the attraction range in the momentum space
is very narrow. The latter condition is violated not only in the GSM, but
also in the initial Timmermanns \textit{et al.} model.

\section{Dynamic atoms-to molecules transformation}

\label{sec:transition}

In this section we consider the transformation of Fermi-atoms into molecules
in the cold Fermi gas under the sweep of the magnetic field. In the
Hamiltonian \eqref{Ham-s-mode} the parameter $\epsilon$ is driven by
magnetic field: $\epsilon = \mu_{B}(H-H_{0})$. At magnetic field sweeping,
the detuning energy $\epsilon $ depends on time. The time-dependent
Hamiltonian regulating the process of atoms-to-molecules transformation
reads:

\begin{equation}
\hat{H}=2\epsilon(t)\S _z+\frac{g}{\sqrt{V}}\left(\b\S _{+}+\hat{b}^{\dagger}%
\S _{-}\right)  \label{eq:HamS}
\end{equation}
It formally coincides with the GSM Hamiltonian \eqref{Ham-spin-global} with
the only difference that the detuning energy $\epsilon$ depends on time.
Even the time-dependent Hamiltonian \eqref{eq:HamS} commutes both with the
operator $Q$ given by \eqref{Sz-conservation}, equivalent to the operator of
the total number of atoms $N$, and with the square of the total spin
operator $S^{2}$ \eqref{total-spin} which both remain the integrals of
motion. The Heisenberg equations of motion are:
\begin{equation}
\hbar\dot{\b}=-i\tilde{g}\S _{-};\qquad \hbar\dot{\S }_{-}=-2i\epsilon(t)\S %
_{-}+2i\tilde{g}\hat{b}^{\dagger}\S _z  \label{eq:motionZ}
\end{equation}
where $\tilde{g} =g/\sqrt{V}$ and dots denote the time derivatives.
Generally these equations are non-linear. However, in the broad-resonance
approximation $S_z = -N_{s}/2$, they become linear. Eliminating $S_{-}$, we
arrive at an ordinary linear differential equation for the operator $\b$:
\begin{equation}
\hbar^2\ddot{\b}+2i\hbar\epsilon(t)\dot{\b}+\Delta^2\b=0  \label{c-equation}
\end{equation}
where $\Delta $ is defined in \eqref{eq:D}.

Equation \eqref{c-equation} becomes the parabolic cylinder equation if $%
\epsilon (t) $ is a linear function of time. In the LZ theory it describes
the evolution of the probability amplitude to find the system in one of its
two states. The value $\Delta$ in the corresponding LZ problem is the matrix
element of transition between the two crossing terms, the so-called LZ gap.

As we already argued, during the magnetic field sweep the Anderson spins
with $p\ll p_{s}$ rotate coherently with almost the same angular velocity
and enhance the effective field acting on the BEC condensate. The spins with
$p>p_{s}$ rotate with substantially different angular velocities and
mutually cancel their contribution to the effective field. Our model
coarsens these features, neglecting the decoherence of spins with $p<p_{s}$
and completely ignoring spins with $p>p_{s}$.

An important conclusion is that strong interaction renormalizes the LZ gap.
The energy scale which appears in perturbation theory is $\Delta ^{\left(
0\right) }=g\sqrt{n}$ \cite{DP}. For a broad resonance $\Delta=g\sqrt{n_s}$
is much larger than $\Delta^{\left( 0\right)}$ and does not depend on the
atomic density. 
%{\color{blue} It is necessary to check the experimental data
%by Regal et al. who stated that Delta depends on n.}

Employing equations \eqref{eq:motionZ}, the general solution of the ordinary
differential equation \eqref{c-equation} reads (further we put $\hbar =1$):
\begin{eqnarray}
\b\left( t\right) =u\left( t,t_{0}\right) \b\left( t_{0}\right) - i\tilde{g}
v\left(t,t_{0}\right) \S _{-}\left( t_{0}\right),&&  \label{general} \\
i\tilde{g}\S _{-}(t)=-\dot{u}\left( t,t_{0}\right)\b\left( t_{0}\right)+ i%
\tilde{g} \dot{v}\left(t,t_{0}\right) \S _{-}\left( t_{0}\right),&&
\label{generalS}
\end{eqnarray}
where $u\left( t,t_{0}\right) $ and $v\left( t,t_{0}\right) $ are standard
solutions of the same equation \eqref{c-equation} satisfying the initial
conditions $u\left( t_{0},t_{0}\right) =1$, $\dot{u}\left(
t_{0},t_{0}\right) =0$ and $v\left( t_{0},t_{0}\right) =0$, $\dot{v}\left(
t_{0},t_{0}\right) =1$. These solutions have the following properties:
\begin{eqnarray}  \label{eq:prop}
&&\!\!\!\!\!\left\vert u\right\vert^{2}\!\!+\! \Delta^{-2}\left\vert \dot{u}%
^{2}\right\vert\! =\! \Delta^{2}\left\vert v\right\vert^{2}\!\! +\!
\left\vert \dot{v}\right\vert^{2}\! =\! \left\vert u\right\vert ^{2}\!\!+\!
\Delta^2\left\vert v\right\vert^{2}\!=\!1;  \label{modulus} \\
&&\dot{u}^{\ast}\dot{v}+\Delta^2u^{\ast}v=0; \quad \dot{u} v - u\dot{v}
=e^{-i\int_{t_0}^t \epsilon (t^{\prime}) dt^{\prime}} .  \label{phase}
\end{eqnarray}
The solution (\ref{general},\ref{generalS}) completely determines the
evolution of the number of molecules $N_m(t)=\langle \hat{b}^{\dagger}\b
\rangle(t)$, the BCS condensate amplitude $F(t)$ defined by equation $%
F^2(t)\equiv\langle \S _{+}\S _{-}\rangle(t)$, and the BCS-BEC coherence
factor $C(t)\equiv\langle \hat{b}^{\dagger} \S _{-}\rangle(t)$. It is
convenient to introduce the intensive values: the molecular (or BEC) density
$n_m(t)=N_m(t)/V$, the BCS condensate density $f(t)=g\sqrt{F(t)}/V$ and the
BEC-BCS correlator density $c(t)=-igC(t)/V^{3/2}$. Then equations of motion
read
\begin{eqnarray}
&&n_m(t)=|u|^2n_{m0}+|v|^2f^2_{0}+2\Re \left( u^{\ast}v c_{0}\right)
\label{eq:A} \\
&&f^2(t)=|\dot{u}|^2 n_{m0}+ |\dot{v}|^2 f^{2}_{0} + 2\Re \left( \dot{u}%
^{\ast}\dot{v} c_{0}\right)  \label{eq:B} \\
&&c(t)=\dot{u}u^{\ast} n_{m0}+\dot{v}v^{\ast} f^2_{0} +u^{\ast}\dot{v} c_{0}
+ \dot{u} v^{\ast} c^{\ast}_{0},  \label{eq:C}
\end{eqnarray}
where we also introduced $n_{m0}=n_{m}(t_{0})$, $f_{0}=f(t_{0})
$, and $c_{0}=c(t_{0})$ -- the initial values of the density of molecules,
the BSC condensate density, and the BEC-BCS correlator density. Using %
\eqref{eq:prop} and summing eqs. \eqref{eq:A} and \eqref{eq:B}, we find:
\begin{eqnarray}  \label{eq:const1}
\Delta^2 n_m(t)+f^2(t)=\mbox{const},
\end{eqnarray}
which is a consequence of the conservation laws. Since for any state $%
F^2(t)>0$, if there are no molecules in the initial state, their number $%
N_m(t)$ can not exceed the value $F^2(t_{0})/N_{s}$ at any time.
 Although equation \eqref{eq:const1} has the same meaning as
equation \eqref{eq:staticConservation} it is more general as it
%More generally equation (\ref{eq:const1})
shows that one of condensates can be extended only  at the expense of another
\emph{even  at a finite sweep rate}.

Below we consider two experimentally most relevant situations: i) only
fermions and no molecules; ii) only molecules and no fermions in the initial
state. In both these cases the initial value $C=\langle \hat{b}^{\dagger} \S %
_{-}\rangle(t_{0})=0$. In the case of no molecules in the initial state, so $%
N_m(t_{0})=0$, the general equations (\ref{eq:A},\ref{eq:C}) simplify to
\begin{equation}  \label{average-t}
N_m(t)= \tilde{g}^2 |v|^{2}F^2(t_0);\quad i\tilde{g} C(t)= -\tilde{g}^2\dot{v%
}v^{\ast}F^2(t_0)
\end{equation}
The evolution of $F(t)$ in this case is determined by \eqref{eq:const1} and %
\eqref{average-t}. Note that the coherence factor $C(t)$ does not remain
zero.

In order to produce a reasonable fraction of molecules it is necessary to
have a large condensate amplitude in the initial state with the only
exception for the equilibrium initial state and adiabatically slow sweep of
magnetic field. Due to the finite gap in the spectrum our model predicts
that in this situation the system will adiabatically follow the ground
state. However, the initial state of the gas in the experiment is reached by
the pumping of the ac electromagnetic field and it is plausible that it is
not in equilibrium.

A strong dependence of the final molecular production on the initial state
(in particular on the value of the initial magnetic field) explains why
different experimenters obtain different fractions of molecules in the final
state even in the adiabatic regime \cite{Strecker2003, Zwierlein2003,
Greiner2003, Regal2003}.

In experiments, which achieved a significant molecular production, the
initial state was indeed close to the FR, whereas the final state was rather
far from the resonance. Thus, in a realistic experimental setup the initial
value of $\epsilon $ is small $|\epsilon_0|\leq\Gamma\epsilon_F\ll \Delta$
and then $\epsilon$ increases linearly with time. In this case one can put $%
t_0=0$, and $\epsilon (t) =\dot{\epsilon}t$. Equation \eqref{c-equation}
turns into the parabolic cylinder equation. Its standard solution $u(t,0)$
has the asymptotic behavior: $|u(\infty,0)|^2=\exp (-\pi \Delta^{2}/2\hbar%
\dot{\epsilon})$. Employing it together with \eqref{eq:prop} and %
\eqref{average-t}, we arrive at the following number of molecules in the
final state:
\begin{equation}
N_m (+\infty )=F^2_0N_{s}^{-1}\left[ 1-\exp \left( -\pi \Delta ^{2}/2\hbar%
\dot{\epsilon}\right) \right]  \label{LZ-answer-bb}
\end{equation}
Eq. \eqref{eq:BCS-epsilon} implies that the maximal possible value of $F^2$
is $N_s N/2$.   %!! I inserted a dot.
%\footnote{It corresponds to the $N/2$ pairs coherently spread over $N_{s}$ states.}.
It corresponds to the complete transformation of atoms into molecules in the
adiabatic regime $\dot{\epsilon}\rightarrow 0$. Equation \eqref{LZ-answer-bb}
looks exactly like the LZ transition probability multiplied by an effective
number of pairs. However, in contrast to phenomenological theories \cite%
{Mies2000,Chwedenczuk2004} and the perturbation theory result \cite{DP}, the
coefficient in front of $1/\dot{\epsilon}$ in the exponent does not depend
on the initial atomic density. This theoretical prediction can be checked
experimentally.
%{\VP In the D. Jin experiments they stated that the exponent depends on density, but as I remember, the data were so scattered that it is difficult to judge whether this conclusion is justified. We should reconsider their data.}
The interaction and many-body effects influence the effective number of
pairs (pre-exponent), which is proportional to the square of the initial BCS
condensate amplitude $F(t_0)$. Finite temperature destroys a fraction of the
initial Cooper pairs and decreases the molecular production.

Finally, we consider an inverse process with no fermions, no BCS condensate
and only the molecular condensate in the initial state: $\langle \b \rangle
( -\infty ) =\sqrt{N/2}$ and sweeping of the magnetic field in the opposite
direction. Then at the end of the sweeping the condensate density is
determined by the LZ value: $\langle \b \rangle ( +\infty ) = \sqrt{N/2}%
\exp(-\pi \Delta ^{2}/ 2\hbar\dot{\epsilon})$, whereas the absolute value of
the BCS condensate amplitude $\langle\S _{-}\rangle $ can be found from the
conservation law \eqref{eq:const1} (for macroscopic condensate amplitude we
can neglect the non-commutativity of $S_{+}$ and $S_{-}$.):
\begin{equation}
\vert \langle \S _{-}\rangle \vert ^{2}=\frac{N_{s}N}{2} \left[ 1-\exp
\left( -\frac{\pi \Delta ^{2}}{\hbar\dot{\epsilon}}\right) \right]
\label{bcs}
\end{equation}
Notice  the factor of $2$ difference in the exponents of \eqref{LZ-answer-bb} and \eqref{bcs}.

An important problem is the reversibility of the process. Eq. %
\eqref{c-equation} for the amplitude $b$ is time-reversible. It means that,
if $b(t)$ is a solution of \eqref{c-equation}  in a time interval $(t_1,t_2)$,
then $b^{\ast}(\bar{t})$ is
also a solution of the same equation with $\epsilon(t)$ replaced
by $\epsilon(\bar{t})$, where $\bar{t}$ denotes the time in the same
interval passed in opposite direction. Unfortunately, it is not clear how
to reverse the phase of the BEC in the experiment. If we simply start with
the same number of molecules in the initial state, which we just before
obtained after passing the FR, it does not guarantee that we reproduce in
the end the state of the Fermi gas from which the molecules were obtained in
the first half-cycle. It is obvious for the non-adiabatic process with not
too large LZ parameter $\Delta^2/\dot{\epsilon}$. In this case to achieve
reversibility we need not only to reproduce correct ratio of number of
molecules to the BCS condensate density, but also their relative phase which
is large and quickly varying value. Even in the adiabatic situation the
reversibility is scarcely reachable for two reasons. First, all levels
except of the ground state have numerous crossings with other levels (see
the previous section) and the adiabatic approximation for them is invalid as
it was indicated for more general situation by Polkovnikov and Gritsev\cite%
{polkovnikov}. But even when we start with the ground state and sweep the
magnetic field adiabatically, the asymptotic value of LZ process is reached
only after very long time, since the amplitudes of "parasite" states decay
slowly, as $1/t$. The real experiment stops at some finite time, and
corrections to the LZ result can be sufficiently large. They are
unpredictable and represent a source of irreversibility in the experiment.

%The quasimolecules have two peculiarities. First, in contrast to the real         %!! This is repetition. Besides, we did not introduce the mane "quasimolecule'
%molecules the quasimolecules have parallel electron spins. Second, they are              in this text.
%unstable: after magnetic field sweep ends if the trap is still maintained,
%the quasimolecules decay. The relaxation time is rather long since fermion
%pair collisions do not produce energy relaxation. The experimental estimate
%for the relaxation time is in the range from milliseconds to seconds.
%{\VP Where it comes from? More details of the relaxation processes are needed.}
%Therefore, it seems quite feasible to switch off the trap before the
%quasimolecules relax and observe the correlations of momenta and spins in
%runaway particles. The prediction of our theory is that the correlation
%prefers opposite velocities and parallel spins in the range of energy up to $%
%\Delta$. The remaining molecules give rise to correlated atoms with opposite
%momenta and spins, providing an alternative opportunity to find their number.

\section{Conclusion}

\label{sec:conclusion}

 First we estimate numerical values of different
constants entering our theory from available experimental data. The numerical
value $g$ can be extracted from the experimental data on the magnetic field
dependence of the scattering length $a$ near the FR \cite{Regal2004} using
the well-known relation: $g=\hbar\sqrt{4\pi(a-a_0)\varepsilon /m}$ ($a_0$ is
the scattering length far from resonance). On the other hand $g$ can be
estimated theoretically as $g\sim \epsilon _{hf}\sqrt{a_{0}^{3}}$, where $%
\epsilon _{hf}$ is the hyperfine energy \cite{LLv3}. Both these estimates
give for $^{40}$K $g\sim 10^{-28}erg\times cm^{3/2}$ and from eqs. %
\eqref{ns-g} and \eqref{eq:D} $\Delta \sim 3\times 10^{-4}K$. However, eq. %
\eqref{eq:D} overestimates $\Delta$ by assuming that the limiting kinetic
energy is $\Delta$ instead of it being much smaller. A more reliable
estimate can be extracted from a comparison of eq. \eqref{LZ-answer-bb} with
experimental data by Regal \textit{et al.} \cite{Regal2003}. The fitting
gives the value $\Delta\sim 10^{-5}K$ for the broad resonance at $B_0=224.21G
$ in $^{40}$K. The cited measurements were performed at the finite
temperature $T\sim T_F/3$,and therefore the corresponding value of $\Delta$
is underestimated in comparison to its zero-temperature value. Thus, a
reasonable estimate for $\Delta$ is between $10^{-5}$K and $10^{-4}$K. In
the cited experiment the magnetic field sweep amplitude was about 12G. It
corresponds to an energy scale of about $10^{-3}K$, larger than $\Delta$.

%{\VP In conclusion the results of the static part must be included. The level
%crossing makes the adiabatic process impossible even at very slow sweep.
%Polkovnikov's article must be cited for general consideration. The dynamics
%conclusions must be supported by the analysis of different relaxation times.
%I am not sure that we should start with the estimate of g. The references
%in the conclusion are copied from our brief article. They must be corrected.
%}

In conclusion, we considered a cooled Fermi-gas in the magnetic field close
the broad Feshbach resonance. We showed that for this strong coupling regime 
simplifying assumptions stemming from the broad resonance conditions allow  %!! I omitted part of this sentence as a tautology and corrected the verb alllows to plural
one to solve both static
and dynamic problems of the BEC-BCS conversion exactly. It was demonstrated
that in this situation the single mode approximation is appropriate. The
neglect of the fermion kinetic energy, being a rough approximation, gives
qualitatively correct physical picture. With these two assumptions we
derived the Global Spin Model Hamiltonian \eqref{Ham-spin-global} and its
Hilbert space for the static case. For the dynamic conversion problem we
solved the Landau-Zener problem for operators.

For the static problems we have found complete spectrum, the number of
molecules and its fluctuations, the amplitude of the BCS condensate and its
correlation with the amplitude of the BEC condensate at different values of
the detuning parameter $\epsilon$ (see equations \eqref{M-G}, %
\eqref{fluctuation}, \eqref{eq:BCS-epsilon}, and \eqref{BCS-BEC}).  %!! I introduced round brackets

One of the most surprising results is the big size of the Cooper pair in the
momentum space and small its size in configurational space. Though we
obtained this result in the framework of the GSM, it will remain in more
precise models since it is a consequence of a large extension of the
hyperfine interaction in momentum space in comparison with the Fermi
momentum. Experimentally this theoretical prediction can be verified by the
time-flight spectroscopy combined with the spin correlation analysis. After
switching off the trap and the driving magnetic field the Cooper pairs
become unstable and decay forming the correlated pairs of particles with
opposite velocities and parallel spins. The relaxation time is rather long
since fermion pair collisions do not produce energy relaxation. The
experimental estimate for the relaxation time is in the range from
milliseconds to seconds.
%{\VP Where it comes from? More details of the relaxation processes are needed.}
Therefore, it seems quite feasible to switch off the trap before the
quasimolecules relax and observe the correlations of momenta and spins in
runaway particles.

For the dynamic problem we have considered molecule formation and
dissociation when the magnetic field is swept across the broad Feshbach
resonance. %It was demonstrated that in this situation the fermion kinetic
%energy is negligible.
The resulting molecular production from initial fermions is described by
LZ-like formulae with a strongly renormalized LZ gap independent of the
initial fermion density. However, the molecular production strongly depends
on the initial value of magnetic field. Though being reversible in
principle, the process is irreversible in practice since the preparation of
the inverse process requires fixing of uncontrollable phases and slowly
decaying corrections to the LZ asymptotics. Another experimentally
verifiable prediction is the independence of the coefficient in front of $1/%
\dot{\epsilon}$ in the LZ exponents for the molecular \eqref{LZ-answer-bb}
and the BCS condensate \eqref{bcs} productions of the initial density of
atoms (molecules).

\acknowledgments

%{\VP Should we rewrite the acknowledgement?}

We acknowledge illuminating discussions with Victor Gurarie, Leo
Radzyhovsky, and A. Polkovnikov. Our thanks due to W. Saslow for attentive
reading of the manuscript and constructive %!!!
suggestions. Ar.A. gratefully acknowledges financial support from Welch
Foundation and NSF Grant PHY-0757992. V.P. is thankful to the Kavli
Institute of %!!!
Theoretical Physics, UCSB, for the hospitality extended to him during the
Workshop on coherence effects in cold gases and strongly correlated systems,
May - June 2007. This work was partly supported by the DOE under the grant
DE-FG02-06ER46278 and KITP NSF Grant No. PHY05-51164.

\bibliography{ColdFermiGas,LL}

\begin{thebibliography}{41}
\expandafter\ifx\csname natexlab\endcsname\relax\def\natexlab#1{#1}\fi
\expandafter\ifx\csname bibnamefont\endcsname\relax
  \def\bibnamefont#1{#1}\fi
\expandafter\ifx\csname bibfnamefont\endcsname\relax
  \def\bibfnamefont#1{#1}\fi
\expandafter\ifx\csname citenamefont\endcsname\relax
  \def\citenamefont#1{#1}\fi
\expandafter\ifx\csname url\endcsname\relax
  \def\url#1{\texttt{#1}}\fi
\expandafter\ifx\csname urlprefix\endcsname\relax\def\urlprefix{URL }\fi
\providecommand{\bibinfo}[2]{#2}
\providecommand{\eprint}[2][]{\url{#2}}

\bibitem[{\citenamefont{Chwedenczuk et~al.}(2004)}]{Chwedenczuk2004}
\bibinfo{author}{\bibfnamefont{J.}~\bibnamefont{Chwedenczuk}}
  \bibnamefont{et~al.}, \bibinfo{journal}{\prl} \textbf{\bibinfo{volume}{93}},
  \bibinfo{pages}{260403} (\bibinfo{year}{2004}).

\bibitem[{\citenamefont{G\'{o}ral et~al.}(2004)}]{Goral2004}
\bibinfo{author}{\bibfnamefont{K.}~\bibnamefont{G\'{o}ral}}
  \bibnamefont{et~al.}, \bibinfo{journal}{Journal of Physics B: Atomic,
  Molecular and Optical Physics} \textbf{\bibinfo{volume}{37}},
  \bibinfo{pages}{3457} (\bibinfo{year}{2004}).

\bibitem[{\citenamefont{Mies et~al.}(2000)\citenamefont{Mies, Tiesinga, and
  Julienne}}]{Mies2000}
\bibinfo{author}{\bibfnamefont{F.~H.} \bibnamefont{Mies}},
  \bibinfo{author}{\bibfnamefont{E.}~\bibnamefont{Tiesinga}}, \bibnamefont{and}
  \bibinfo{author}{\bibfnamefont{P.~S.} \bibnamefont{Julienne}},
  \bibinfo{journal}{\pra} \textbf{\bibinfo{volume}{61}},
  \bibinfo{pages}{022721} (\bibinfo{year}{2000}).

\bibitem[{\citenamefont{S.~Inouye et~al.}(1998)}]{Inouye1998}
\bibinfo{author}{\bibfnamefont{S.}~\bibnamefont{S.~Inouye}}
  \bibnamefont{et~al.}, \bibinfo{journal}{Nature}
  \textbf{\bibinfo{volume}{392}}, \bibinfo{pages}{151} (\bibinfo{year}{1998}).

\bibitem[{\citenamefont{Stwalley}(1976)}]{Stwalley1976}
\bibinfo{author}{\bibfnamefont{W.~C.} \bibnamefont{Stwalley}},
  \bibinfo{journal}{\prl} \textbf{\bibinfo{volume}{37}}, \bibinfo{pages}{1628}
  (\bibinfo{year}{1976}).

\bibitem[{\citenamefont{Tiesinga et~al.}(1993)\citenamefont{Tiesinga, Verhaar,
  and Stoof}}]{Tiesinga1993}
\bibinfo{author}{\bibfnamefont{E.}~\bibnamefont{Tiesinga}},
  \bibinfo{author}{\bibfnamefont{B.~J.} \bibnamefont{Verhaar}},
  \bibnamefont{and} \bibinfo{author}{\bibfnamefont{H.~T.~C.}
  \bibnamefont{Stoof}}, \bibinfo{journal}{\pra} \textbf{\bibinfo{volume}{47}},
  \bibinfo{pages}{4114} (\bibinfo{year}{1993}).

\bibitem[{\citenamefont{Timmermans et~al.}(1999)\citenamefont{Timmermans,
  Tommasini, Hussein, and Kerman}}]{Timmermans1999}
\bibinfo{author}{\bibfnamefont{E.}~\bibnamefont{Timmermans}},
  \bibinfo{author}{\bibfnamefont{P.}~\bibnamefont{Tommasini}},
  \bibinfo{author}{\bibfnamefont{M.}~\bibnamefont{Hussein}}, \bibnamefont{and}
  \bibinfo{author}{\bibfnamefont{A.}~\bibnamefont{Kerman}},
  \bibinfo{journal}{Physics Reports} \textbf{\bibinfo{volume}{315}},
  \bibinfo{pages}{199 } (\bibinfo{year}{1999}), ISSN \bibinfo{issn}{0370-1573}.

\bibitem[{\citenamefont{Chin et~al.}(2003)}]{Chin2003}
\bibinfo{author}{\bibfnamefont{C.}~\bibnamefont{Chin}} \bibnamefont{et~al.},
  \bibinfo{journal}{\prl} \textbf{\bibinfo{volume}{90}},
  \bibinfo{pages}{033201} (\bibinfo{year}{2003}).

\bibitem[{\citenamefont{Cubizolles et~al.}(2003)}]{Cubizolles2003}
\bibinfo{author}{\bibfnamefont{J.}~\bibnamefont{Cubizolles}}
  \bibnamefont{et~al.}, \bibinfo{journal}{\prl} \textbf{\bibinfo{volume}{91}},
  \bibinfo{pages}{240401} (\bibinfo{year}{2003}).

\bibitem[{\citenamefont{Donley et~al.}(2002)}]{Donley2002}
\bibinfo{author}{\bibfnamefont{E.~A.} \bibnamefont{Donley}}
  \bibnamefont{et~al.}, \bibinfo{journal}{Nature}
  \textbf{\bibinfo{volume}{417}}, \bibinfo{pages}{529} (\bibinfo{year}{2002}).

\bibitem[{\citenamefont{Durr et~al.}(2004)}]{Durr2004}
\bibinfo{author}{\bibfnamefont{S.}~\bibnamefont{Durr}} \bibnamefont{et~al.},
  \bibinfo{journal}{\prl} \textbf{\bibinfo{volume}{92}},
  \bibinfo{pages}{020406} (\bibinfo{year}{2004}).

\bibitem[{\citenamefont{Herbig et~al.}(2003)}]{Herbig2003}
\bibinfo{author}{\bibfnamefont{J.}~\bibnamefont{Herbig}} \bibnamefont{et~al.},
  \bibinfo{journal}{Science} \textbf{\bibinfo{volume}{301}},
  \bibinfo{pages}{1510} (\bibinfo{year}{2003}).

\bibitem[{\citenamefont{Jochim et~al.}(2003)}]{Jochim2003}
\bibinfo{author}{\bibfnamefont{S.}~\bibnamefont{Jochim}} \bibnamefont{et~al.},
  \bibinfo{journal}{Science} \textbf{\bibinfo{volume}{302}},
  \bibinfo{pages}{2101} (\bibinfo{year}{2003}).

\bibitem[{\citenamefont{Regal et~al.}(2003)}]{Regal2003}
\bibinfo{author}{\bibfnamefont{C.}~\bibnamefont{Regal}} \bibnamefont{et~al.},
  \bibinfo{journal}{Nature} \textbf{\bibinfo{volume}{424}}, \bibinfo{pages}{47}
  (\bibinfo{year}{2003}).

\bibitem[{\citenamefont{Strecker et~al.}(2003)\citenamefont{Strecker,
  Partridge, and Hulet}}]{Strecker2003}
\bibinfo{author}{\bibfnamefont{K.~E.} \bibnamefont{Strecker}},
  \bibinfo{author}{\bibfnamefont{G.~B.} \bibnamefont{Partridge}},
  \bibnamefont{and} \bibinfo{author}{\bibfnamefont{R.~G.} \bibnamefont{Hulet}},
  \bibinfo{journal}{\prl} \textbf{\bibinfo{volume}{91}},
  \bibinfo{pages}{080406} (\bibinfo{year}{2003}).

\bibitem[{\citenamefont{Greiner et~al.}(2003)\citenamefont{Greiner, Regal, and
  Jin}}]{Greiner2003}
\bibinfo{author}{\bibfnamefont{M.}~\bibnamefont{Greiner}},
  \bibinfo{author}{\bibfnamefont{C.~A.} \bibnamefont{Regal}}, \bibnamefont{and}
  \bibinfo{author}{\bibfnamefont{D.~S.} \bibnamefont{Jin}},
  \bibinfo{journal}{Nature} \textbf{\bibinfo{volume}{426}},
  \bibinfo{pages}{537} (\bibinfo{year}{2003}).

\bibitem[{\citenamefont{Hodby et~al.}(2005)}]{Hodby2005}
\bibinfo{author}{\bibfnamefont{E.}~\bibnamefont{Hodby}} \bibnamefont{et~al.},
  \bibinfo{journal}{\prl} \textbf{\bibinfo{volume}{94}},
  \bibinfo{pages}{120402} (\bibinfo{year}{2005}).

\bibitem[{\citenamefont{Zwierlein et~al.}(2003)}]{Zwierlein2003}
\bibinfo{author}{\bibfnamefont{M.~W.} \bibnamefont{Zwierlein}}
  \bibnamefont{et~al.}, \bibinfo{journal}{\prl} \textbf{\bibinfo{volume}{91}},
  \bibinfo{pages}{250401} (\bibinfo{year}{2003}).

\bibitem[{\citenamefont{Claussen et~al.}(2002)}]{Claussen2002}
\bibinfo{author}{\bibfnamefont{N.~R.} \bibnamefont{Claussen}}
  \bibnamefont{et~al.}, \bibinfo{journal}{\prl} \textbf{\bibinfo{volume}{89}},
  \bibinfo{pages}{010401} (\bibinfo{year}{2002}).

\bibitem[{\citenamefont{Eagles}(1969)}]{Eagles1969}
\bibinfo{author}{\bibfnamefont{D.~M.} \bibnamefont{Eagles}},
  \bibinfo{journal}{Phys. Rev.} \textbf{\bibinfo{volume}{186}},
  \bibinfo{pages}{456} (\bibinfo{year}{1969}).

\bibitem[{\citenamefont{Leggett}(1980)}]{Leggett1980}
\bibinfo{author}{\bibfnamefont{A.}~\bibnamefont{Leggett}}, in
  \emph{\bibinfo{booktitle}{Modern trends in the theory of condensed matter}}
  (\bibinfo{publisher}{Springer-Verlag, Berlin}, \bibinfo{year}{1980}), pp.
  \bibinfo{pages}{13--27}.

\bibitem[{\citenamefont{Nozieres and Schmitt-Rink}(1985)}]{Nozieres1985}
\bibinfo{author}{\bibfnamefont{P.}~\bibnamefont{Nozieres}} \bibnamefont{and}
  \bibinfo{author}{\bibfnamefont{S.}~\bibnamefont{Schmitt-Rink}},
  \bibinfo{journal}{Journal of Low Temperature Physics}
  \textbf{\bibinfo{volume}{59}}, \bibinfo{pages}{195} (\bibinfo{year}{1985}).

\bibitem[{\citenamefont{Chen et~al.}(2005)\citenamefont{Chen, Stajic, Tan, and
  Levin}}]{Levin2005}
\bibinfo{author}{\bibfnamefont{Q.}~\bibnamefont{Chen}},
  \bibinfo{author}{\bibfnamefont{J.}~\bibnamefont{Stajic}},
  \bibinfo{author}{\bibfnamefont{S.}~\bibnamefont{Tan}}, \bibnamefont{and}
  \bibinfo{author}{\bibfnamefont{K.}~\bibnamefont{Levin}},
  \bibinfo{journal}{Physics Reports} \textbf{\bibinfo{volume}{412}},
  \bibinfo{pages}{1} (\bibinfo{year}{2005}).

\bibitem[{\citenamefont{S\'a~de Melo et~al.}(1993)\citenamefont{S\'a~de Melo,
  Randeria, and Engelbrecht}}]{deMelo1993}
\bibinfo{author}{\bibfnamefont{C.~A.~R.} \bibnamefont{S\'a~de Melo}},
  \bibinfo{author}{\bibfnamefont{M.}~\bibnamefont{Randeria}}, \bibnamefont{and}
  \bibinfo{author}{\bibfnamefont{J.~R.} \bibnamefont{Engelbrecht}},
  \bibinfo{journal}{Phys. Rev. Lett.} \textbf{\bibinfo{volume}{71}},
  \bibinfo{pages}{3202} (\bibinfo{year}{1993}).

\bibitem[{\citenamefont{Dobrescu and Pokrovsky}(2006)}]{DP}
\bibinfo{author}{\bibfnamefont{B.~E.} \bibnamefont{Dobrescu}} \bibnamefont{and}
  \bibinfo{author}{\bibfnamefont{V.~L.} \bibnamefont{Pokrovsky}},
  \bibinfo{journal}{Phys. Lett.} \textbf{\bibinfo{volume}{A350}},
  \bibinfo{pages}{154} (\bibinfo{year}{2006}).

\bibitem[{\citenamefont{Timmermans et~al.}(2001)}]{Timmermanns2001}
\bibinfo{author}{\bibfnamefont{E.}~\bibnamefont{Timmermans}}
  \bibnamefont{et~al.}, \bibinfo{journal}{Phys. Lett. A}
  \textbf{\bibinfo{volume}{285}}, \bibinfo{pages}{228} (\bibinfo{year}{2001}).

\bibitem[{\citenamefont{Javanainen et~al.}(2004)}]{Javanainen2004}
\bibinfo{author}{\bibfnamefont{J.}~\bibnamefont{Javanainen}}
  \bibnamefont{et~al.}, \bibinfo{journal}{\prl} \textbf{\bibinfo{volume}{92}},
  \bibinfo{pages}{200402} (\bibinfo{year}{2004}).

\bibitem[{\citenamefont{Tikhonenkov et~al.}(2006)\citenamefont{Tikhonenkov,
  Pazy, and Vardi}}]{Tikhonenkov2006-OC}
\bibinfo{author}{\bibfnamefont{I.}~\bibnamefont{Tikhonenkov}},
  \bibinfo{author}{\bibfnamefont{E.}~\bibnamefont{Pazy}}, \bibnamefont{and}
  \bibinfo{author}{\bibfnamefont{A.}~\bibnamefont{Vardi}},
  \bibinfo{journal}{Optics Communications} \textbf{\bibinfo{volume}{264}},
  \bibinfo{pages}{321} (\bibinfo{year}{2006}).

\bibitem[{\citenamefont{Williams et~al.}(2004)}]{Williams2004}
\bibinfo{author}{\bibfnamefont{J.~E.} \bibnamefont{Williams}}
  \bibnamefont{et~al.}, \bibinfo{journal}{J. Phys. B}
  \textbf{\bibinfo{volume}{37}}, \bibinfo{pages}{L351} (\bibinfo{year}{2004}).

\bibitem[{\citenamefont{Gurarie and Radzihovsky}(2007)}]{Gurarie2007}
\bibinfo{author}{\bibfnamefont{V.}~\bibnamefont{Gurarie}} \bibnamefont{and}
  \bibinfo{author}{\bibfnamefont{L.}~\bibnamefont{Radzihovsky}},
  \bibinfo{journal}{Annals of Physics} \textbf{\bibinfo{volume}{322}},
  \bibinfo{pages}{2} (\bibinfo{year}{2007}).

\bibitem[{\citenamefont{Dicke}(1954)}]{dicke}
\bibinfo{author}{\bibfnamefont{R.~H.} \bibnamefont{Dicke}},
  \bibinfo{journal}{Phys. Rev.} \textbf{\bibinfo{volume}{93}},
  \bibinfo{pages}{99} (\bibinfo{year}{1954}).

\bibitem[{\citenamefont{Sun et~al.}(2008)\citenamefont{Sun, Abanov, and
  Pokrovsky}}]{SAP}
\bibinfo{author}{\bibfnamefont{D.}~\bibnamefont{Sun}},
  \bibinfo{author}{\bibfnamefont{A.}~\bibnamefont{Abanov}}, \bibnamefont{and}
  \bibinfo{author}{\bibfnamefont{V.~L.} \bibnamefont{Pokrovsky}},
  \bibinfo{journal}{Europhysics Letters} \textbf{\bibinfo{volume}{83}},
  \bibinfo{pages}{16003} (\bibinfo{year}{2008}).

\bibitem[{\citenamefont{Landau and Lifshitz}(2002)}]{LLv3}
\bibinfo{author}{\bibfnamefont{L.~D.} \bibnamefont{Landau}} \bibnamefont{and}
  \bibinfo{author}{\bibfnamefont{E.~M.} \bibnamefont{Lifshitz}},
  \emph{\bibinfo{title}{Course of Theoretical Physics, Quantum Mechanics}},
  vol.~\bibinfo{volume}{3} (\bibinfo{publisher}{Butterworth-Heinenann},
  \bibinfo{address}{Linacre House, Jordan Hill, Oxford OX2 8DP, England},
  \bibinfo{year}{2002}), \bibinfo{edition}{3rd} ed., \bibinfo{note}{translated
  from Russian by J. B. Sykes and J. S. Bell}.

\bibitem[{\citenamefont{Lifshitz and Pitaevskii}(2002)}]{LLv9}
\bibinfo{author}{\bibfnamefont{E.~M.} \bibnamefont{Lifshitz}} \bibnamefont{and}
  \bibinfo{author}{\bibfnamefont{L.~P.} \bibnamefont{Pitaevskii}},
  \emph{\bibinfo{title}{Course of Theoretical Physics, Statistical Physics,
  Part 2}}, vol.~\bibinfo{volume}{9}
  (\bibinfo{publisher}{Butterworth-Heinenann}, \bibinfo{address}{Linacre House,
  Jordan Hill, Oxford OX2 8DP, England}, \bibinfo{year}{2002}),
  \bibinfo{edition}{1st} ed., \bibinfo{note}{translated from Russian by J. B.
  Sykes and M. J. Kearsley}.

\bibitem[{\citenamefont{Kagan and Svistunov}(1994)}]{Kagan-Svist}
\bibinfo{author}{\bibfnamefont{Y.}~\bibnamefont{Kagan}} \bibnamefont{and}
  \bibinfo{author}{\bibfnamefont{B.~V.} \bibnamefont{Svistunov}},
  \bibinfo{journal}{Sov. Phys. JETP} \textbf{\bibinfo{volume}{78}},
  \bibinfo{pages}{187} (\bibinfo{year}{1994}).

\bibitem[{\citenamefont{Kagan et~al.}(1992)\citenamefont{Kagan, Svistunov, and
  Shlyapnikov}}]{Kagan-Svist-Shlyap}
\bibinfo{author}{\bibfnamefont{Y.}~\bibnamefont{Kagan}},
  \bibinfo{author}{\bibfnamefont{B.}~\bibnamefont{Svistunov}},
  \bibnamefont{and}
  \bibinfo{author}{\bibfnamefont{G.}~\bibnamefont{Shlyapnikov}},
  \bibinfo{journal}{Sov. Phys. JETP} \textbf{\bibinfo{volume}{75}},
  \bibinfo{pages}{387} (\bibinfo{year}{1992}).

\bibitem[{\citenamefont{Gardiner et~al.}(1997)\citenamefont{Gardiner, Zoller,
  Ballagh, and Davis}}]{Gardiner1}
\bibinfo{author}{\bibfnamefont{C.~W.} \bibnamefont{Gardiner}},
  \bibinfo{author}{\bibfnamefont{P.}~\bibnamefont{Zoller}},
  \bibinfo{author}{\bibfnamefont{R.~J.} \bibnamefont{Ballagh}},
  \bibnamefont{and} \bibinfo{author}{\bibfnamefont{M.~J.} \bibnamefont{Davis}},
  \bibinfo{journal}{Phys. Rev. Lett.} \textbf{\bibinfo{volume}{79}},
  \bibinfo{pages}{1793} (\bibinfo{year}{1997}).

\bibitem[{\citenamefont{Gardiner et~al.}(1998)\citenamefont{Gardiner, Lee,
  Ballagh, Davis, and Zoller}}]{Gardiner2}
\bibinfo{author}{\bibfnamefont{C.~W.} \bibnamefont{Gardiner}},
  \bibinfo{author}{\bibfnamefont{M.~D.} \bibnamefont{Lee}},
  \bibinfo{author}{\bibfnamefont{R.~J.} \bibnamefont{Ballagh}},
  \bibinfo{author}{\bibfnamefont{M.~J.} \bibnamefont{Davis}}, \bibnamefont{and}
  \bibinfo{author}{\bibfnamefont{P.}~\bibnamefont{Zoller}},
  \bibinfo{journal}{Phys. Rev. Lett.} \textbf{\bibinfo{volume}{81}},
  \bibinfo{pages}{5266} (\bibinfo{year}{1998}).

\bibitem[{\citenamefont{Anderson}(1958)}]{AndersonTrans}
\bibinfo{author}{\bibfnamefont{P.~W.} \bibnamefont{Anderson}},
  \bibinfo{journal}{Phys. Rev.} \textbf{\bibinfo{volume}{112}},
  \bibinfo{pages}{1900} (\bibinfo{year}{1958}).

\bibitem[{\citenamefont{Polkovnikov and Gritsev}(2008)}]{polkovnikov}
\bibinfo{author}{\bibfnamefont{A.}~\bibnamefont{Polkovnikov}} \bibnamefont{and}
  \bibinfo{author}{\bibfnamefont{V.}~\bibnamefont{Gritsev}},
  \bibinfo{journal}{Nature Physics} \textbf{\bibinfo{volume}{4}},
  \bibinfo{pages}{477} (\bibinfo{year}{2008}).

\bibitem[{\citenamefont{Regal et~al.}(2004)\citenamefont{Regal, Greiner, and
  Jin}}]{Regal2004}
\bibinfo{author}{\bibfnamefont{C.~A.} \bibnamefont{Regal}},
  \bibinfo{author}{\bibfnamefont{M.}~\bibnamefont{Greiner}}, \bibnamefont{and}
  \bibinfo{author}{\bibfnamefont{D.~S.} \bibnamefont{Jin}},
  \bibinfo{journal}{\prl} \textbf{\bibinfo{volume}{92}},
  \bibinfo{pages}{040403} (\bibinfo{year}{2004}).

\end{thebibliography}

\end{document}